\def\papertitle{What you hear is what you see:\\ Audio quality from Image Quality Metrics}
\def\paperauthorA{Tashi Namgyal}
\def\paperauthorB{Alexander Hepburn}
\def\paperauthorC{Valero Laparra}
\def\paperauthorD{Raul Santos-Rodriguez}
\def\paperauthorE{Jesus Malo}

\documentclass[twoside,a4paper]{article}
\usepackage{etoolbox}
\usepackage{dafx_23}
\usepackage{amsmath,amssymb,amsfonts,amsthm}
\usepackage{euscript}
\usepackage[T1]{fontenc}
\usepackage[utf8]{inputenc}
\usepackage{ifpdf}
\usepackage[english]{babel}
\usepackage{caption}
\usepackage{subfig} 
\usepackage{color}
\usepackage{booktabs}
\usepackage{todonotes}

\input glyphtounicode
\ninept

\newcounter{numauth}
\newcounter{listcnt}
\newcommand\authcnt[1]{\ifdefined#1 \stepcounter{numauth} \fi}

\newcommand\addauth[1]{
\ifdefined#1 
\stepcounter{listcnt}
\ifnum \value{listcnt}<\value{numauth}
\appto\authorslist{, #1}
\else
\appto\authorslist{~and~#1}
\fi
\fi}
\authcnt{\paperauthorB}
\authcnt{\paperauthorC}
\authcnt{\paperauthorD}
\authcnt{\paperauthorE}
\authcnt{\paperauthorF}
\authcnt{\paperauthorG}
\authcnt{\paperauthorH}
\authcnt{\paperauthorI}
\authcnt{\paperauthorJ}
\def\authorslist{\paperauthorA}
\addauth{\paperauthorB}
\addauth{\paperauthorC}
\addauth{\paperauthorD}
\addauth{\paperauthorE}
\addauth{\paperauthorF}
\addauth{\paperauthorG}
\addauth{\paperauthorH}
\addauth{\paperauthorI}
\addauth{\paperauthorJ}

\usepackage{times}

\newif\ifpdf
\ifx\pdfoutput\relax
\else
   \ifcase\pdfoutput
      \pdffalse
   \else
      \pdftrue
\fi

\ifpdf 
  \usepackage[pdftex,
    pdftitle={\papertitle},
    pdfauthor={\authorslist},
    pdfsubject={Proceedings of the 26th International Conference on Digital Audio Effects (DAFx23)},
    colorlinks=false, 
    bookmarksnumbered, 
    pdfstartview=XYZ 
  ]{hyperref}
\else 
  \usepackage[dvips]{epsfig,graphicx}
  \usepackage[dvips,
    pdftitle={\papertitle},
    pdfauthor={\authorslist},
    pdfsubject={Proceedings of the 26th International Conference on Digital Audio Effects (DAFx23)},
    colorlinks=false, 
    bookmarksnumbered, 
    pdfstartview=XYZ 
  ]{hyperref}
\fi
\usepackage[hypcap=true]{caption}
\title{\papertitle}

\twoaffiliations{
\paperauthorA, \, \paperauthorB, \, \paperauthorD  \thanks{\vspace{-3mm}}}
{{Intelligent Systems Lab} \\ University of Bristol\\ Bristol, UK\\
{\tt \href{tashi.namgyal@bristol.ac.uk}{tashi.namgyal@bristol.ac.uk} \href{alex.hepburn@bristol.ac.uk}{alex.hepburn@bristol.ac.uk} \href{enrsr@bristol.ac.uk}{enrsr@bristol.ac.uk}}
}
{\paperauthorC, \, \paperauthorE}
{\href{https://ipl.uv.es/}{Image Processing Lab} \\ Universitat de Valencia \\ Valencia, Spain \\ {\tt \href{valero.laparra@uv.es}{valero.laparra@uv.es} \\ \href{jesus.malo@uv.es}{jesus.malo@uv.es}}
}

\usepackage{todonotes}
\begin{document}
\ifpdf 
  \DeclareGraphicsExtensions{.png,.jpg,.pdf}
\else  
  \DeclareGraphicsExtensions{.eps}
\fi

\maketitle

\begin{abstract}

In this study, we investigate the feasibility of utilizing state-of-the-art perceptual image metrics for evaluating audio signals by representing them as spectrograms. The encouraging outcome of the proposed approach is based on the similarity between the neural mechanisms in the auditory and visual pathways. Furthermore, we customise one of the metrics which has a psychoacoustically plausible architecture to account for the peculiarities of sound signals. We evaluate the effectiveness of our proposed metric and several baseline metrics using a music dataset, with promising results in terms of the correlation between the metrics and the perceived quality of audio as rated by human evaluators.
\end{abstract}

\section{Introduction}
\label{sec:intro}

Perceptual assessment of the quality of audio signals has been explored to varying degrees for different kinds of audio content. Whilst there exist several tools to understand speech quality~\cite{gamper2019intrusive}, the evaluation of music is rarely explored and comes in the form of software hidden behind commercial licences~\cite{thiede2000peaq}. More generally, practitioners rely either on traditional physical measures of the audio signal, e.g., signal-to-noise ratio (SNR), or more recent deep learning-based metrics that involve noninterpretable models to capture statistics of the degradation~\cite{hilmkil2020perceiving}. The picture is quite different in the visual modality, where many more perceptual models have been developed over the years for these purposes -- and well-curated datasets are readily available~\cite{ponomarenko2015image, zhang2018unreasonable}.

It is well-known that the auditory and visual processing pathways share similar attributes. For example, \textit{divisive normalisation}, a form of local gain control, is a well explored phenomenon that is encountered when studying neurons in the brain~\cite{carandini2012normalization, schwartz2001natural}.  Specifically in vision, divisive normalisation has been shown to factorise the probability density function of natural images~\cite{malo2010psychophysically}. In audio the same phenomenon has been shown to minimise the dependencies between between natural sound stimuli responses to filters of certain frequencies~\cite{schwartz2001natural}. 
Other behaviours such as signal adaptation can also be observed in both modalities~\cite{willmore2023adaptation}. Many of these ideas form the basis of
the design of image quality metrics, but, as they are also observed in auditory statistics or psychophysical tests, we argue they should be included in the design of audio quality metrics.

The algorithmic parallelism and interaction between neural pathways implies that audio signals can alter the perception of visual stimuli~\cite{Shams00}, and examples of this (audio-to-vision) correlation are present even in pop music~\cite{Mertens06}. Here we take the opposite (vision-to-audio) approach: \emph{what you hear is what you see}.

We draw inspiration from state-of-the-art image quality metrics to bridge the gap with their audio counterparts, which are not so successful at predicting perceived quality, for example, when evaluating neural audio synthesis~\cite{vinay2022evaluating}. Although raw audio takes a very different form to images, well-studied transformations can be used to align the two modalities. For example, spectrograms represent audio signals using image-like 2D matrices, where each column represents a time window and each row is a frequency band. As such, spectrograms encode the audio signal similar to wavelet decompositions that are often used in image metrics~\cite{malo2010psychophysically,laparra2010metric}. We can then use these representations to exploit the literature on image quality metrics (IQMs) to estimate audio quality. Importantly, whilst the structure and semantics of spectrograms are different to images, the underlying principles are similar, e.g. the importance of amplitude (brightness) and local differences
(contrast). 

The paper is organised as follows: firstly, we show that popular IQMs can outperform metrics specifically designed for audio. Secondly we show that fine-tuning a traditional IQM based on divisive normalisation, which is also seen in auditory processing, can further improve results. We also provide the intuition behind what this tuned metric is capturing about the properties of audio.


\section{Quality Metrics}
Quality metrics aim to replicate the distance between two examples perceived by a human. This usually involves projecting the raw data to a perceptually meaningful space and computing a distance, or computing and comparing statistical descriptors of the examples. Below we will detail a number of audio and image quality metrics used throughout the paper.

\subsection{Image Quality Metrics}
Traditional IQMs fall into two categories; \emph{structural similarity}, comparing descriptions of image statistics, and \emph{visibility of errors}, which aims to measure how visible the distortions are to humans. Multi-Scale Structural SIMilarity (MS-SSIM)~\cite{wang2003multiscale} is based on the former and compares three descriptors 
(luminance, contrast and structure) at various scales. Normalised Laplacian Pyramid Distance (NLPD)~\cite{laparra2016}, based on the visibility of errors, is inspired by the biological processing in the visual system. Coincidentally, this processing is also present in the auditory system and we will use this to fine-tune NLPD to audio (sec.~\ref{sec:NLPD}). 

\subsection{Audio Quality Metrics}
Audio quality metrics have typically been designed for evaluating audio coding and source separation artifacts~\cite{torcoli2021objective}. 
Here, we compare three recent metrics. 
Fréchet Audio Distance (FAD)~\cite{kilgour2019frechet} is a reference-free metric for evaluating generated audio based on the Fréchet Inception Distance (FID) commonly used in images~\cite{heusel2017gans}. FAD uses embeddings from the VGGish model~\cite{hershey2017cnn} to measure the distance between previously learned clean studio quality music and a given audio clip. Virtual Speech Quality Objective Listener (ViSQOL)~\cite{chinen2020visqol} is a full-reference metric based on the Neural Similarity Measure (NSIM)~\cite{hines2012speech} between spectrograms. NSIM is similar to SSIM, using the luminance and structure terms but dropping the contrast term. Additionally it uses a support vector regression model to map the NSIM scores more closely to Mean Opinion Scores. The discriminator output of a Generative Adversarial Network (GAN) can also be used to predict perceptual ratings~\cite{hilmkil2020perceiving}.

\section{Normalised Laplacian Pyramid Distance}
\label{sec:NLPD}

NLPD is our example case for adapting existing image metrics to audio. The Laplacian Pyramid is well known in image coding~\cite{burt1983laplacian}. The signal is encoded by applying a low-pass filter and then subtracting this from the original image multiple times at various scales, creating low entropy versions of the signal. The Normalised Laplacian Pyramid (NLP) extends this with a local normalisation on the output of each pyramid level~\cite{laparra2016}. These two steps are similar to the early stages of the visual and auditory systems where linear filtering and local normalisation are present~\cite{schwartz2001natural,laparra2010metric,willmore2023adaptation}. The distance in this new domain is referred to as NLPD~\cite{laparra2016,Laparra17}, correlates well with human perception, and reduces redundancy 
in agreement with the efficient coding hypothesis~\cite{malo2010psychophysically}. 
\begin{figure}[t]
    \centering
    \includegraphics[width=.85\columnwidth]{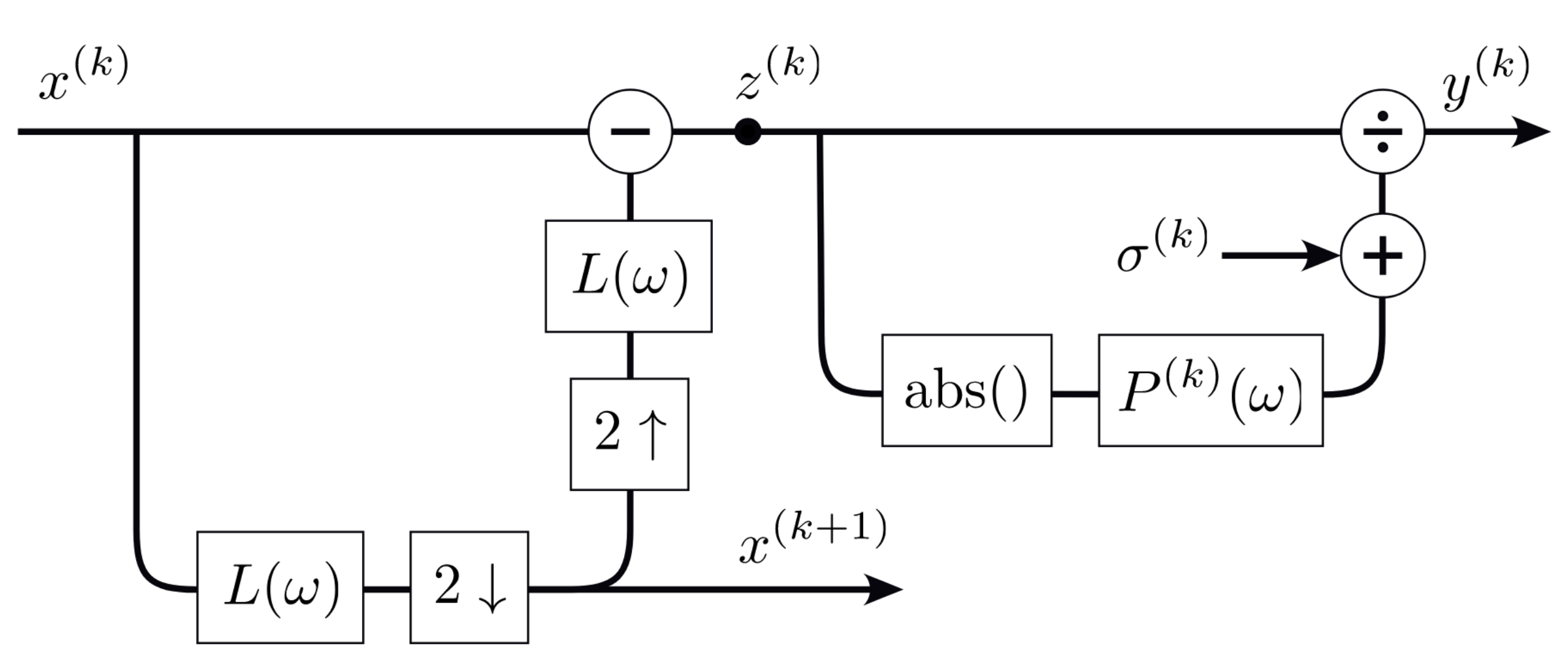}
    \vspace{-0.25cm}
    \caption{{\it Architecture for one stage $k$ of the Normalised Laplacian Pyramid model, where $x^{(k)}$ is the input at stage $k$, $L(\omega)$ is a convolution with a low-pass filter, $[2\downarrow]$ is a downsample of factor two, $[2\uparrow]$ is an upsample of factor two, $x^{(k+1)}$ is the input image at stage $(k+1)$, $P^{(k)}(\omega)$ is s scale-specific filter for normalising the image with respect to the local amplitude, $\sigma^{(k)}$ is scale-specific constant and $y^{(k)}$ is the output at scale $k$. Figure taken from \cite{laparra2016}.}}
    \label{fig:nlap-arch}
\end{figure}

An overview of the architecture is detailed in Fig. \ref{fig:nlap-arch}. 
Given two images, $x_1$ and $x_2$, we compute the outputs $y^{(k)}_1$ and $y^{(k)}_2$ at every stage of the pyramid $k$, and sum the differences:
\begin{equation}
    \textrm{NLPD}(x_1,x_2) = \frac{1}{N}\sum^N_{k=1}\frac{1}{\sqrt{N_s^{(k)}}}||y_1^{(k)} - y_2^{(k)}||_{_2}
\end{equation} where $N$ is the number of stages in the pyramid, and $N_s^{(k)}$ is the number of coefficients at stage $k$. 

\section{Experiments}

\subsection{Data}

We use the Perceived Music Quality Dataset from~\cite{hilmkil2020perceiving}. It consists of 4-second audio clips across 13 genres, with 5 songs per genre and 3 clips per song, totalling 195 reference clips. These reference clips are degraded in four ways: waveshape distortion, low pass filtering, limiting and additive noise, resulting in 975 clips. We divide this into an 80-20 train-test split, in which the test set contains all 3 clips for the last song in each genre. Each clip has an associated perceptual quality rating on a scale from 1 to 5 
These ratings were gathered using Amazon Mechanical Turk using a no-reference paradigm. Each clip was rated by at least 5 participants and the median value was taken. 

For the SSIM, NLPD, and Mean Square Error (MSE) metrics the audio clips are downmixed into mono and converted into mel spectrograms. The audio is downsampled from 48kHz to 16050Hz with 512 mel-bands, a window size of 2048, and a hop-length of 64, resulting in spectrograms of size $512\times1024$. For NLPD we use 6 pyramidal layers, with inputs being halved in size for each layer down to $16\times32$. The SSIM ratings are calculated using Pytorch MS-SSIM\footnote{\footnotesize{{https://github.com/VainF/pytorch-msssim}}}. For ViSQOL and FAD, the clips are downmixed into mono and converted from 32-bit to 16-bit WAV files. Ratings are calculated using the ViSQOL\footnote{\footnotesize{{https://github.com/google/visqol}}} and FAD\footnote{\footnotesize{\href{https://github.com/google-research/google-research/tree/master/frechet\_audio\_distance}{{--/google-research/google-research/tree/master/frechet\_audio\_distance}}}} packages. 


\begin{table}[htb]
\centering
\caption{{\it Spearman correlation between human ratings and various metrics. NLPD~\cite{laparra2016} and (MS-)SSIM~\cite{wang2003multiscale} are image quality metrics, whilst ViSQOL~\cite{chinen2020visqol}, FAD~\cite{kilgour2019frechet} and GAN~\cite{hilmkil2020perceiving} are audio quality metrics. We report the correlation for each degradation type separately as well as for all degradations simultaneously. GAN results are taken from the respective paper~\cite{hilmkil2020perceiving}.}}
\vspace{-0.23cm}
\begin{tabular}{@{}lccccc@{}}
\toprule
\multicolumn{1}{l}{Metric} &
  \multicolumn{1}{c}{Waveshape} &
  \multicolumn{1}{c}{Lowpass} &
  Limiter &
  \multicolumn{1}{c}{Noise} &
  \multicolumn{1}{c}{All} \\ \midrule
MSE        & \textbf{0.469} & -0.049 & \textbf{0.378} & 0.641 & 0.483 \\
NLPD       & 0.468 & 0.012 & 0.339 & \textbf{0.681} & 0.633 \\
SSIM       & -0.450 & -0.175 & -0.356 & -0.629 & \textbf{-0.656} \\ 
MS-SSIM    & -0.468 & -0.045 & -0.323 & -0.654 & -0.648 \\
ViSQOL     & -0.142 & 0.191 & -0.316 & -0.629 & -0.232 \\
FAD        & 0.386 & -0.083 & 0.316 & 0.550 & 0.593 \\
GAN*       & 0.349 & \textbf{0.222} & 0.120 & 0.359 & 0.426 \\ \bottomrule
\end{tabular}

\label{tab:results1}
\end{table}

\subsection{NLPD Optimisation}
To fine-tune NLPD to audio we optimise the filters $P^{(k)}(\omega)$ and the constant $\sigma^{(k)}$ in the divisive normalisation stages. There are two possible methods; statistically~\cite{laparra2016} or perceptually~\cite{laparra2010metric}. We use 5x5 filters in both cases.

Optimising statistically consists of calculating average pixel values of the band-passed spectrograms $z$ separately for each layer $k$. The divisive normalisation filters are learned as weights $p_j$ that transform the weighted sum of pixel values in the neighbourhood surrounding each pixel to approximate the centre pixel, $j$:
\begin{equation} \label{eq:fc}
    f_C^{(k)}\left(\mathbf{z}_{N_i}\right) = \mathbf{\sigma}^{(k)} + \sum_{j \in N_i} p_j^{(k)} \left|z_j^{(k)}\right|
\end{equation}
where $N_i$ defines the neighbourhood (filter size) to be considered. The constant $\sigma^{(k)}$ is the mean absolute value of $z$ for each layer:
\begin{equation} \label{eq:sig}
    \mathbf{\sigma}^{(k)}=\frac{1}{N_s^{(k)}}\sum_{i=1}^{N_s^{(k)}}\left|z_i^{(k)}\right|
\end{equation}
where $N_s$ is the number of coefficients at stage $k$, i.e. dimensions of $z$. The weights are optimised with Eq.~\ref{eq:p}. We optimise over the reference spectrograms contained in the training set only, using ADAM optimiser, learning rate 0.01, batch size of 1 for 10 epochs. 
\begin{equation} \label{eq:p}
    \mathbf{\hat{p}}^{(k)} = \underset{\mathbf{p}}{\mathrm{argmin}} \sum_{i=1}^{N_s^{(k)}}\left(\left|z_i^{(k)}\right|-f_C\left(\mathbf{z}_{N_i}^{(k)}\right)\right)^2
\end{equation}
Optimising perceptually consists of maximising the Pearson's correlation between the NLPD and the human ratings of each reference audio clip and a degraded version of the clip. The filters are initialised to be the image NLPD values, and $\sigma^{(k)}$ is initialised with Eq~.\ref{eq:sig}. We use ADAM optimiser to maximise the Pearson correlation with a learning rate of 0.001 for 100 epochs, where each batch only contains one degradation. 
We use Pearson as the training objective instead of Spearman's, assuming approximately linear rankings, as the sorting operation has undefined gradients.

\begin{table}[htb]
\centering
\caption{{\it Spearman correlations for variations of the NLPD. Original: filters fit statistically to natural images~\cite{laparra2016}. No DN: NLP with no divisive normalisation stage. $P(\omega)=1$: divisive normalisation filters are all ones. Statistical: filters optimised to predict the center pixel given its neighbours. Perceptual: model optimised to maximise correlation with human ratings.}}
\vspace{-0.25cm}
\begin{tabular}{@{}lccccc@{}}
\toprule
\multicolumn{1}{l}{Metric} & \multicolumn{1}{c}{Waveshape}   & \multicolumn{1}{c}{Lowpass} & \multicolumn{1}{c}{Limiter}     & \multicolumn{1}{c}{Noise}       & \multicolumn{1}{c}{All} \\ \midrule
Original                   & \textbf{0.468} & 0.012                                                                  & 0.339                           & \textbf{0.681} & 0.633                                                                  \\
No DN                      & 0.412                           & -0.052                                                                 & 0.336                           & 0.670                           & 0.617                                                                  \\
$P(\omega)=1$              & 0.457                           & -0.022                                                                 & \textbf{0.380} & 0.669                           & 0.629                                                                  \\
Statistical                & 0.432                           & -0.033                                                                 & 0.356                           & 0.660                           & 0.619                                                                  \\
Perceptual                 & 0.430                           & \textbf{0.035}                                        & 0.347                           & 0.637                           & \textbf{0.643}                                        \\ \bottomrule
\end{tabular}

\label{tab:res2}
\end{table}


\begin{figure*}[h]
    \centering
    \includegraphics[width=.7\textwidth]{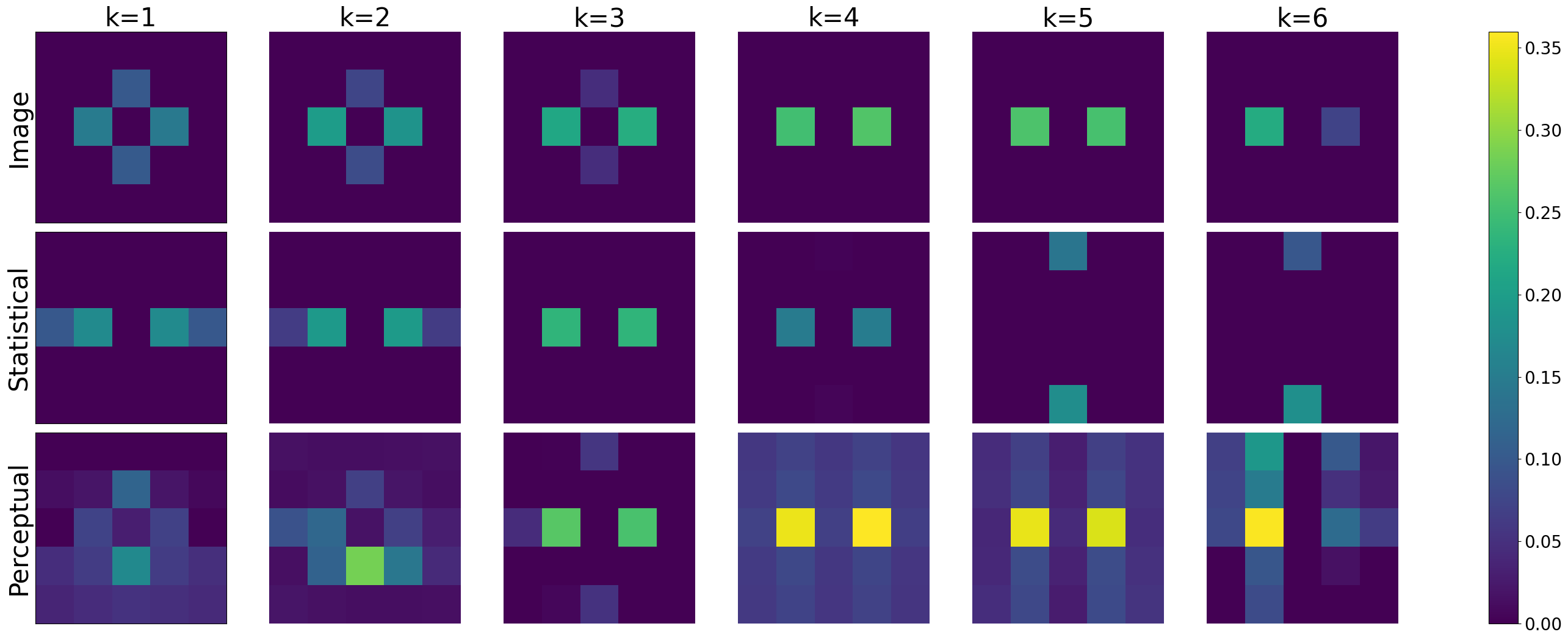}
    \vspace{-0.25cm}
    \caption{{\it Divisive normalisation filters learnt different optimisation strategies. Each column is a different layer in the pyramid, $k$. The top row (image) is the implementation of NLPD used for images, the second row (statistical) is statistically fit to audio spectrograms (Eq.~{\ref{eq:p}}), and the final row (perceptual) are filters resulting from fitting NLPD to perceptual judgements on audio.}}
    \label{fig:perceptual_filters}
\end{figure*}

\section{Results and Discussion}

\subsection{Main findings}
Table~\ref{tab:results1} shows correlations between humans and perceptual quality metrics. Surprisingly, IQMs perform better than AQMs for all degradations other than the low pass filter. However, the limiter and low pass filter had much weaker p-values
so these correlations could be due to chance. 
We think this is partly because the amount of degradation applied in those cases 
was not high enough compared to the waveshaping and noise degradations. This is indicated by the degraded audio being judged as better quality than the reference in some pairs. 

Table~\ref{tab:res2} shows results for adapting NLPD to audio using five different divisive normalisation strategies: 
1. using filters from NLPD optimised statistically on natural images (as in Table \ref{tab:results1}), 
2. with no divisive normalisation, 
3. setting the filters in the divisive normalisation to one (for equal contribution of all the neighbours), 
4. fitting the model statistically on spectrograms (where no perceptual information is used), and 
5. maximising the correlation between spectrograms and the opinion of humans. 
The perceptually trained divisive normalisation has the highest correlation when all degradations are tested simultaneously, and other strategies trained on spectrograms increase correlation for the low pass and limiter degradations. For waveshape and noise, the forms of divisive normalisation using spectrograms decrease the correlation compared to training on natural images. 
The relationship between the degradations tested and the form of divisive normalisation used could be further explored as this process may effectively be reducing certain degradations, i.e. enhancing the signal.

Fig.~\ref{fig:perceptual_filters} shows the learned divisive normalisation filters at different layers of the NLPD for 3 optimisation strategies. For the first four layers, the statistical spectrogram model focuses almost exclusively within the central frequency band, particularly at the time steps immediately before and after the central bin. This is similar to later layers of the model fit to natural images but different from early layers, where both directions are important. The later layers of the statistical spectrogram model look across bands but only within a single timestep in a manner completely unlike the image model. This may reflect the pattern of repeating harmonics in spectrogram signals. This may only be captured at later layers as early layers have a higher resolution, i.e. there are more frequency bands between harmonics. 
Using larger filters at early layers may help to capture this. 
The fact that the model only uses the central timestep at later layers may reflect the way that later layers are effectively averaging across longer time windows. As such, the signal will vary less smoothly between time bins and so will be less predictive of the central value. 
Larger time filters may start to capture rhythmic information. In contrast, the perceptual spectrogram filters consider both time and frequency simultaneously, like early layers of the image model, with layers exhibiting more smoothing behaviour in general. This may indicate that perceptually trained models may be better at capturing degradations that effect lower energy regions than statistical models. 


\subsection{Further Work}

We have identified a need for a greater number of publicly available datasets of perceptual quality in audio with a larger variety of sounds and degradation types. The scores in the GAN and ViSQOL papers were collected according to the ITU-T P.800 recommendation (for telephone conversations). However, according to ITU-R BS.1534-1 this proved insufficient for evaluating audio signals of broadcast quality. Instead the "MUlti Stimulus test with Hidden Reference and Anchor" is the recommended grading procedure, as is used in~\cite{vinay2022evaluating}. The non-adaptive psychophysical Two Alternate Forced Choice (2AFC) paradigm, as used in IQMs~\cite{ponomarenko2015image, zhang2018unreasonable} would also be suitable. Preliminary tests should be performed to ensure the range of degradation is similar across degradation types. Tests should also scale degradation amounts to avoid improving the perceptual quality above the reference. A task training procedure or better task descriptions could also improve rating quality, as according to \cite{hilmkil2020perceiving}, participants were asked "How do you rate the audio quality of this music segment?" where "quality" is left largely up to participants to interpret. 
Such a dataset would allow for a more reliable comparison with contemporary~\cite{manocha2021cdpam,gupta2022parameter} and future AQMs. We also plan to investigate how divisive normalisation may be better tailored to audio, such as by using separate filters for time and frequency. We intend to use these metrics as a loss function in generative modelling, so that such models generate audio samples that sound more realistic with fewer perceived distortions. We also want to investigate the degree to which navigating through latent spaces of models trained with perceptual metrics aligns with human expectations of how the generated audio should change. We believe this should help with explainability, trust and control over the outputs of generative audio models.



\section{Acknowledgments}
TN is supported by the UKRI AI CDT (EP/S022937/1). AH and RSR are supported by UKRI Turing AI Fellowship EP/V024817/1. VL and JM are supported by MINCEO and ERDF grants PID2020-118071GB-I00, DPI2017-89867-C2-2-R and GV/2021/074.


\nocite{*}
\bibliographystyle{IEEEbib}
\bibliography{camera_ready} 

\begin{thebibliography}{10}

\bibitem{gamper2019intrusive}
H.~Gamper et~al.,
\newblock ``Intrusive and non-intrusive perceptual speech quality using
  {CNNs},''
\newblock {\em IEEE WASPAA}, pp. 85--9, 2019.

\bibitem{thiede2000peaq}
T.~Thiede et~al.,
\newblock ``{PEAQ}-the {ITU} standard for measurement of perceived audio
  quality,''
\newblock {\em J. Audio-Eng. Soc.}, vol. 48, no. 1/2, pp. 3--29, 2000.

\bibitem{hilmkil2020perceiving}
A.~Hilmkil, C.~Thom{\'e}, and A.~Arpteg,
\newblock ``Perceiving music quality with {GAN}s,''
\newblock {\em arXiv:2006.06287}, 2020.

\bibitem{ponomarenko2015image}
N.~Ponomarenko et~al.,
\newblock ``{TID}2013: Peculiarities, results and perspectives,''
\newblock {\em Sig.Proc.Im.Comm.}, vol. 30, pp. 57--77, 2015.

\bibitem{zhang2018unreasonable}
R.~Zhang et~al.,
\newblock ``The unreasonable effectiveness of deep features as a perceptual
  metric,''
\newblock {\em IEEE CVPR}, pp. 586--95, 2018.

\bibitem{carandini2012normalization}
M.~Carandini and D.~J. Heeger,
\newblock ``Normalization as a canonical neural computation,''
\newblock {\em Nat.Rev.Neurosci.}, vol. 13, pp. 51--62, 2012.

\bibitem{schwartz2001natural}
O.~Schwartz and E.~P. Simoncelli,
\newblock ``Natural signal statistics and sensory gain control,''
\newblock {\em Nat.Neurosci.}, vol. 4, pp. 819--825, 2001.

\bibitem{malo2010psychophysically}
J.~Malo and V.~Laparra,
\newblock ``Psychophysically tuned divisive normalization factorizes the {PDF}
  of natural images,''
\newblock {\em Neural Comput.}, vol. 22, no. 12, pp. 3179--3206, 2010.

\bibitem{willmore2023adaptation}
B.~Willmore and A.~King,
\newblock ``Adaptation in auditory processing,''
\newblock {\em Physiol.Rev.}, vol. 103, no. 2, pp. 1025--1058, 2023.

\bibitem{Shams00}
L.~Shams, Y.~Kamitani, and S.~Shimojo,
\newblock ``What you see is what you hear,''
\newblock {\em Nature}, vol. 408, no. 12, pp. 788, 2000.

\bibitem{Mertens06}
W.~Mertens,
\newblock ``What you see is what you hear. {U}sura {R}ecords. {B}elgium,''
  2006.

\bibitem{vinay2022evaluating}
A.~Vinay and A.~Lerch,
\newblock ``Evaluating generative audio systems and their metrics,''
\newblock in {\em ISMIR}, 2022.

\bibitem{laparra2010metric}
V.~Laparra et~al.,
\newblock ``Divisive normalization image quality metric revisited,''
\newblock {\em JOSA A}, vol. 27, no. 4, pp. 852--64, 2010.

\bibitem{wang2003multiscale}
Z.~Wang, E.~P. Simoncelli, and A.~C. Bovik,
\newblock ``Multiscale structural similarity for image quality,''
\newblock in {\em 37th {IEEE} Asilomar Conf. Sig. Syst. Comp.}, 2003, vol.~2,
  pp. 1398--1402.

\bibitem{laparra2016}
V.~Laparra, J.~Ball{\'e}, A.~Berardino, and E.~P. Simoncelli,
\newblock ``Perceptual image quality assessment using a normalized laplacian
  pyramid,''
\newblock {\em Electr.Imag.}, vol. 2016, no. 16, pp. 1--6, 2016.

\bibitem{torcoli2021objective}
M.~Torcoli et~al.,
\newblock ``Objective measures of perceptual audio quality reviewed,''
\newblock {\em IEEE/ACM Trans. Audio, Speech, Lang. Proc.}, vol. 29, pp.
  1530--1541, 2021.

\bibitem{kilgour2019frechet}
K.~Kilgour et~al.,
\newblock ``Frechet audio distance: A reference-free metric for evaluating
  music enhancem.,''
\newblock {\em InterSpeech}, 2019.

\bibitem{heusel2017gans}
M.~Heusel et~al.,
\newblock ``{GANs} trained by a two time-scale update rule converge to a local
  {N}ash equilibrium,''
\newblock in {\em NeurIPS}, 2017, p. 6626–6637.

\bibitem{hershey2017cnn}
S.~Hershey et~al.,
\newblock ``{CNN} architectures for large-scale audio classification,''
\newblock in {\em IEEE ICASSP}, 2017, pp. 131--135.

\bibitem{chinen2020visqol}
M.~Chinen et~al.,
\newblock ``{ViSQOL}: An open source production-ready object. speech and audio
  metric,''
\newblock {\em IEEE QoMEX}, 2020.

\bibitem{hines2012speech}
A.~Hines and Harte. N,
\newblock ``Speech intelligibility prediction using a neurogram similarity
  index measure,''
\newblock {\em Speech Comm.}, vol. 54, no. 2, pp. 306 – 320, 2012.

\bibitem{burt1983laplacian}
P.~Burt and E.~Adelson,
\newblock ``Laplacian pyramid as a compact image code,''
\newblock {\em IEEE Trans.Comm.}, vol. 31, pp. 532--540, 1983.

\bibitem{Laparra17}
V.~Laparra et~al.,
\newblock ``Perceptually optimized image rendering,''
\newblock {\em JOSA A}, vol. 34, 2017.

\bibitem{manocha2021cdpam}
P.~Manocha et~al.,
\newblock ``{CDPAM}: Contrastive learning for perceptual audio similarity,''
\newblock in {\em ICASSP}, 2021.

\bibitem{gupta2022parameter}
C.~Gupta et~al.,
\newblock ``Parameter sensitivity of deep-feature based evaluation metrics for
  audio textures,''
\newblock in {\em ISMIR}, 2022.

\end{thebibliography}


\end{document}